# Accurately Determining Defect Ionization Energy in Low-Dimensional Semiconductors: Charge Corrected Jellium Model

Guojun Zhu[†], Xin-Gao Gong[†‡*] and Ji-Hui Yang[†*]

[†]Department of Physics, Key Laboratory for Computational Science (MOE), State Key Laboratory of Surface Physics, Fudan University, Shanghai 200433, China

[‡] Collaborative Innovation Center of Advanced Microstructures, Nanjing 210093, Jiangsu, China

**Abstract:**

**Determination of defect ionization energy in low-dimensional semiconductors has been a long-standing unsolved problem in first-principles defect calculations because the commonly used methods based on jellium model introduce an unphysical charge density uniformly distributed in the material and vacuum regions, causing the well-known divergence issue of charged defect formation energies. Here in this work, by considering the physical process of defect ionization, we propose a charge correction method based on jellium model to replace the unphysical jellium background charge density with the band edge charge density to deal with charged defects. We demonstrate that, our method is physically meaningful, quantitatively accurate and technically simple to determine the defect ionization energies, thus solving the long-standing problem in defect calculations. Our proposed method can be applied to any dimensional semiconductors.**

Understanding defect behaviors has been lying in the heart of semiconductor applications, as defects play many important roles in utilizing semiconductor technologies through creating additional levels in the host materials[1-4]. For example, by ionization at finite temperatures, defects with shallow levels close to band edges can provide free carriers and thus give life to semiconductors[4, 5]. On the other hand, defects with deep levels in the band gaps can act as carrier killers by trapping free carriers and assisting the recombination of electrons and holes[6, 7]. Besides, defects can form color centers in semiconductors, which can be used for quantum information and computation[8, 9]. Among all the defect-related properties, defect formation energy and ionization energy (IE) (also known as transition energy level referenced to band edges) are the most important two key quantities: the former determines defect concentration under equilibrium conditions; the latter, defined as the energy cost to get ionized, determines the ability of a defect to provide carriers [2, 9-12]. To determine defect IE, both formation energies of neutral and charged defects should be known, which can be obtained from first-principles defect calculations based on density-functional theory (DFT). During the past decades, defect calculations have been widely performed for three-dimensional (3D) systems[11, 13-17], which have provided guidance of defect control and engineering to boost the performance and efficiency of many devices[11, 18, 19].

Usually, a supercell structure model along with periodic boundary conditions (PBCs) is adopted and a defect and its periodic images are created during the simulation[20-22]. For a neutral defect under dilute approximation, the formation energy can be accurately obtained using a sufficiently large supercell to eliminate the image interactions. For a charged defect, $q$ electrons are removed (added) from (to) the defect and so do its periodic images. The long-range Coulomb interactions between the defect and its periodic images would induce a divergence of the total energy [23]. To remove such divergence, a so-called jellium model is routinely used with homogenous background charge added to the whole supercell space to neutralize the supercell, get its total energy and thus the formation energy of a charged defect. To eliminate the effect of jellium charge on the defect formation energy, the following argument is often adopted: as the supercell size increases, the jellium charge density will go to zero and the charged defect formation energy will converge with supercell sizes. Once converged formation energies of neutral and charged defects are known, converged defect IE can be calculated. During the past decades, this method has achieved great success to understand defect behaviors in 3D semiconductors [24-28].

Recently, low-dimensional (LD) semiconductor systems have attracted more and more research interests, especially after many two-dimensional (2D) semiconductors, including monolayer BN[29-31], MoS$_2$[32, 33] and phosphorene[34, 35], are fabricated and demonstrated with novel fantastic properties for future electronic and optoelectronic applications. Consequently, the defect behaviors in low-dimensional systems, including 2D monolayers and surfaces, and one-dimensional (1D) nanoribbons, nanowires, and nanotubes, are becoming more and more urgent to be explored. During the past decades, people have tried to do so following the same treatment as that used for 3D semiconductors. Particularly, for a charged defect, a homogenous background charge is added to the whole space of the slab model including the vacuum region[36, 37]. While formation energy of a neutral defect can easily converge by increasing the sizes of the supercell and vacuum regions, formation energy of a charged defect is difficult or never to converge by increasing the size of the vacuum region, which is a well-known issue due to the Coulomb interactions between the charge in the vacuum and that in material regions. To solve the convergence problem, various methods have been tried, such as confining the background charge into a given region, introducing a neutralizing charge by using pseudo-atoms, posteriori corrections to fix the potential at cell boundaries[38-41], etc. While these methods might have solved the convergence issue to some extent, the physical justification is unclarified and the technical feasibility is also kind of complex, i.e., requiring additional artificial parameters or relatively expensive calculations.

Recently, several new methods have been developed to avoid artificial parameters. Based on the conventional jellium model, Wang et. al. derived an analytical form of defect IE as a function of supercell lattice parameters $L_x$, $L_y$ and $L_z$ for 2D semiconductors. They proposed a extrapolation method to obtain converged values of defect IEs[37, 42]. Unfortunately, this method might have problems in describing acceptor defects, resulting from the fact that the vacuum level decreases linearly with the distance away from 2D material planes due to the jellium charge in the vacuum region, which could lead to unphysical charge transfers from materials to the vacuum. Alternatively, Wu, Zhang, and Pantelides (WZP) considered the physical process of defect ionization, in which electrons or holes will be removed from the neutral defect state and excited to the conduction bands or valance bands. Under defect dilute condition, the excited electrons or holes will occupy the conduction band minimum (CBM) or the valance band maximum (VBM) after thermodynamic equilibrium. Based on this process, they proposed to simulate the charged defect by constraining the electron occupation numbers at the defect state and the band edges[24]. By doing this, the defect is charged but the whole supercell is still neutral. Consequently, the total energy of such charged defect state has no divergence issue. However, Deng and Wei commented that in current defect calculations using small supercells, one cannot find the exact VBM or CBM in a defective supercell and artificially choosing a perturbed state just above or below the defect level can cause significant errors if the perturbed state is very different from the real VBM or CBM [25]. Besides, according to the comment by Deng and Wei, any exact state in an infinitely large supercell could become a perturbated state in a small supercell. Consequently, the exact defect state may also not be found in small defective supercells especially if the defect state strongly couples to other states. As current methods all have disadvantages, new methods of studying defect IEs in LD semiconductors are strongly desired.

In this paper, by considering the difference of the charge density in the jellium model and the real charge density for an ionized defect, we propose a charge correction method to deal with charged defects in LD semiconductors. The main idea is to replace the unphysical jellium background charge in the jellium model by the band edge charge (VBM for acceptors or CBM for donors). We show that, the total electronic charge density in our method is exactly the same as the real charge density for an ionized defect, ensuring that our method is physically meaningful. The advantages of our method are as follows. First, the unphysical jellium background charge distributed in the whole space of a charged defective supercell is replaced by the band edge charge distributed only within LD semiconductors in our method, so there is no divergence issue. Second, there is no tunable parameters or no need to artificially identify the defect states in our method and thus possible artificial errors could be avoided. By studying defects in typical 2D materials such as BN, MoS$_2$, and black phosphorous monolayers, we show that compared to our method, other methods based on the jellium model have systematically overestimated defect IEs for LD semiconductors due to the overestimation of Coulomb energy between the background charge and the material charge. Free of any tunable model parameters, our charge correction method based on jellium model provides a physically meaningful, quantitively accurate and technically simple method to deal with charged defects adapted not only to LD but also to 3D semiconductors.

**RESULTS AND DISCUSSION**

**Charge Corrected Jellium Model**. We start the development of our method from the physical process of defect ionization. In the following, we take a shallow acceptor (denoted as $\alpha$) as an example, as shown in Fig. 1. The donor case is given in the Supplemental Materials (Fig. S1). Before ionization, the acceptor state, which is just above the VBM, is neutral with one electron and one hole (see Fig. 1a). All the states below the acceptor state are fully occupied by electrons at $T = 0$. Assume the total number of electrons in the system is $N$. Apparently, the whole system is a ground state and the total energy, denoted as $E^N(\alpha, 0)$, can be easily obtained from the ground state DFT calculations. After ionization, the defect accepts one electron from the valance band, leaving one hole behind. Under thermo-dynamical equilibrium and defect dilute condition, the hole finally will be relaxed to the VBM state, as shown in Fig 1b. The ionized defect state is now an excited state of the $N$-electron system, as shown in Fig. 1b. Once we know the total energy of the excited state [denoted as $\tilde{E}^N(\alpha, -1)$], the defect IE can then be calculated as $IE = \tilde{E}^N(\alpha, -1) - E^N(\alpha, 0)$ according to its definition.

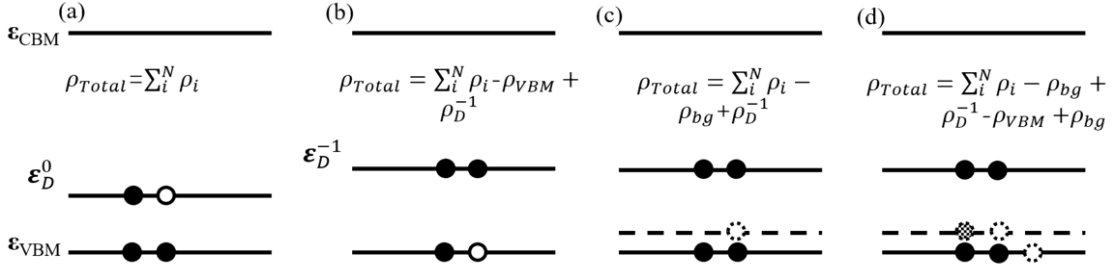

Figure 1. Diagrams to show electron occupations for an acceptor in an infinitely large supercell at (a) the neutral state, (b) the charged ionized state, (c) the charged state in the jellium model framework, and (d) the charge state in our charge correction method. The total charge density is given with $\rho_i$, $\rho_{VBM}$, $\rho_{bg}$ and $\rho_D^{-1}$ standing for charge density in the $i$th band, VBM, background and charged defect level. The electrons and holes are represented by solid and hollow circles, respectively. Solid and dashed lines (circles) stand for real and virtual states (carriers), respectively.

To calculate $\tilde{E}^N(\alpha, -1)$, reasonable approximations are needed. In the defect dilute limit, i.e., in an infinitely large supercell, it's a good approximation that the band edges are not affected by the existence of a defect, which means that if we add one electron back to the VBM, the contribution of this electron to the total energy of the supercell is just the VBM energy $\varepsilon_{VBM}$. Once the electron is added back to the VBM, all the states below the defect state are fully occupied again, that is, the supercell now is a ground state of $(N + 1)$-electron system (as shown in Fig. 1c). Therefore, $\tilde{E}^N(\alpha, -1)$ can be approximated as $E^{N+1}(\alpha, -1) - \varepsilon_{VBM}$ and defect IE can be calculated as $IE = E^{N+1}(\alpha, -1) - E^N(\alpha, 0) - \varepsilon_{VBM}$. This formula has been widely used to calculate defect IEs in the past decades for 3D semiconductors.

Note that, the charged defective supercell in Fig. 1c has $(N + 1)$ electrons and $Ne$ positive charge on the nuclei. To keep the whole system neutral and obtain converged $E^{N+1}(\alpha, -1)$, jellium model is routinely adopted with a background charge uniformly distributed in the whole space of the charged defective supercell [43]. In a 3D system, the jellium charge can be seen as occupying a virtual state with an eigen energy of the Fermi level of the system (Fig. 1c). However, physically the jellium charge should not exist for an ionized defective system. To eliminate the effect of jellium charge, one common method is to increase the supercell size [44]. When the jellium charge density approaches zero, converged formation energy of a charged defect and thus defect IE can be obtained. This argument was the foundation stone in first-principles calculations for charged defects and it worked well for 3D semiconductors. However, when the conventional jellium model meets LD semiconductors, the jellium charge is distributed in the whole supercell space including material and vacuum regions, leading to the well-known divergence of formation energies for charged defects as well as defect IEs. Consequently, one has to look back the physical reasonability of the defect IE calculation methods based on jellium model.

As we know, the total energy is a functional of charge density in DFT. Therefore, if the charge density in a system is unphysical, the total energy and the post physical quantities might also be incorrect. In the jellium model framework for an $(N+1)$-electron system, the charge has the following contributions: $Ne$ positive charge from the nuclei, $(N+1)e$ negative charge from $N+1$ electrons, and $1e$ positive charge from the jellium background. Compared to the real charge density of the ionized defect state (Fig. 1b), one can see that, two parts of the charge in the jellium model are unphysical: the $1e$ positive jellium background charge and the $1e$ negative charge due to the electron added to the VBM. If we replace the unphysical jellium background charge in the jellium model framework by the band edge charge (VBM for acceptors or CBM for donors), i.e., through a charge correction of $\Delta\rho = \rho_{bg} - \rho_{VBM}$ (see Fig. 1d), the electronic charge density in the $(N+1)$-electron supercell restores back to the real charge density of the ionized defect state in Fig. 1b, which has clear physical meanings. Using the corrected charge density, the total energy of $(N+1)$-electron supercell $E_{corr}^{N+1}(\alpha,-1)$ can be calculated. Now under the defect dilute limit, $\tilde{E}^N(\alpha,-1)$ can be approximated as $E_{corr}^{N+1}(\alpha,-1) - \varepsilon_{VBM}$ and the defect IE can be calculated as $IE = E_{corr}^{N+1}(\alpha,-1) - E^N(\alpha,0) - \varepsilon_{VBM}$.

Here we discuss several advantages of our charge correction method to calculate defect IE. First, the uniformly distributed background charge in the jellium model is replaced by the band edge state charge, which is localized within the materials. Consequently, there is no divergence issue for LD semiconductors. Second, after the charge correction, the charge density represents the real charge density for the ionized defect, ensuring our method physically meaningful. Third, the charge density correction $\Delta\rho$ is a constant distribution and doesn't involve any defect states. Therefore, there is no necessary to identify the defect state, thus avoiding possible artificial errors due to using incorrect defect states.

**Calculated defect properties**. First, we demonstrate that using the physically justified charge correction method based on jellium model, converged charge defect formation energies and defect IEs with respect to the sizes of vacuum regions are achieved for LD semiconductors. We take defects in the h-BN monolayer as examples and consider nitrogen vacancy ($V_N$) and carbon substituting nitrogen ($C_N$) as typical donor and acceptor defects, respectively. By fixing the lateral sizes of defective supercells as $12 \times 12 \times 1$ of the primitive cells, Figs. 2a and 2b show the calculated formation energies of charged defects of $V_N^+$ and $C_N^-$ as functions of vacuum thickness. Clearly, in the conventional method based on jellium model, the formation energies of $V_N^+$ and $C_N^-$ increase almost linearly with the length of vacuum regions, in agreement with previous reports [42]. Instead, in our charge correction method, both the formation energies of $V_N^+$ and $C_N^-$ don't change with the vacuum thickness (Figs. 2a and 2b). This can be understood as follows. In the conventional method, the uniformly distributed negative (positive) background charge for a charged donor (acceptor) increases (decreases) the electrostatic potentials in the vacuum regions (Figs. 2c and 2d). For acceptor defects, if the electrostatic potentials drop too much to make the vacuum level lower than the states in the material regions, electrons will transfer from material regions to the vacuum [42], causing the deviation of defect formation energies from linear increase with vacuum thickness (see Fig. S3). In our method, there is no net background charge in the vacuum regions and thus the electron electrostatic potentials are rather flat (see Fig. 2b), leading to the unchanged formation energies for charged defects with respect to vacuum thickness. Consequently, our method gives converged defect IEs, as shown in Fig. 2e.

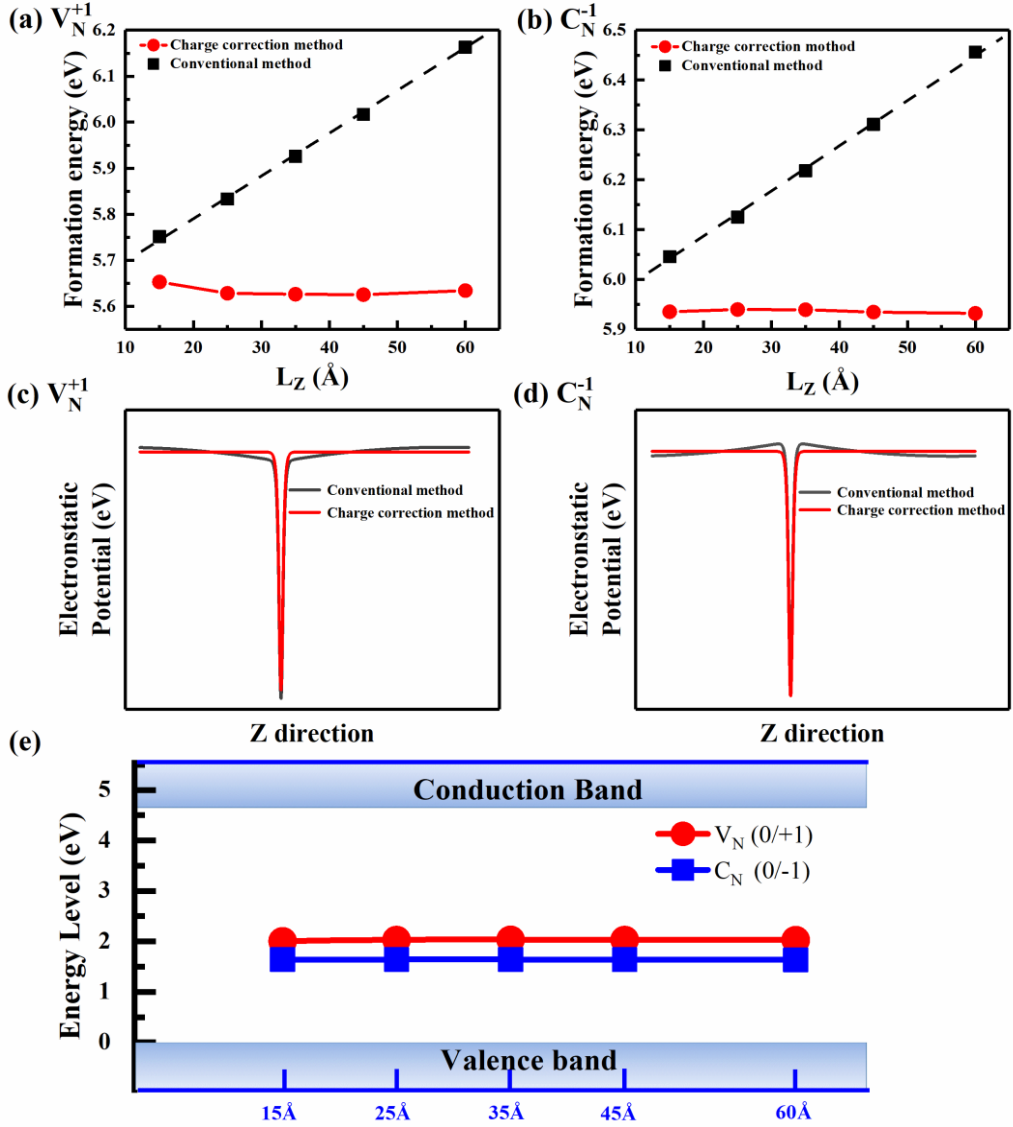

Figure 2. Calculated defect properties of two dimensional BN with the conventional and charge corrected jellium model. (a) and (b) Formation energies of $V_N^+$ and $C_N^-$ dependence on vacuum lengths. (c) and (d) Electrostatic potentials along the vacuum direction of $V_N^+$ and $C_N^-$ supercells with the lengths along vacuum directions fixed at 60 Å. (e) Defect IEs of $V_N^+$ and $C_N^-$ as functions of vacuum lengths. The lateral sizes of all supercells are fixed as $12 \times 12$ of the primitive cells. Note that, the (0/-) level of $C_N^-$ is equal to its defect IE while the (0/+) level of $V_N^+$ is equal to the bandgap value of BN minus the defect IE of $V_N^+$. The dashes lines are added for guiding eyes.

Next, we consider the effects of long-range Coulomb interactions between charged defects and the corresponding images on the defect IEs due to the finite lateral supercell sizes. By fixing the vacuum thickness as 15Å, we gradually increase the lateral sizes of defect supercells from $3 \times 3$ to $15 \times 15$ of the primitive cells. As can be seen in Fig. 3, the calculated defect IEs using our method have already been converged within 0.1 eV using a $12 \times 12$ supercell, which is a typical supercell size in modern defect calculations with affordable computational costs. Combining the results in both Fig. 2 and Fig. 3, we can conclude that our method, based on the physical process of defect ionization and thus physically justified, indeed eliminates the divergence problem in the conventional method based on the jellium model.

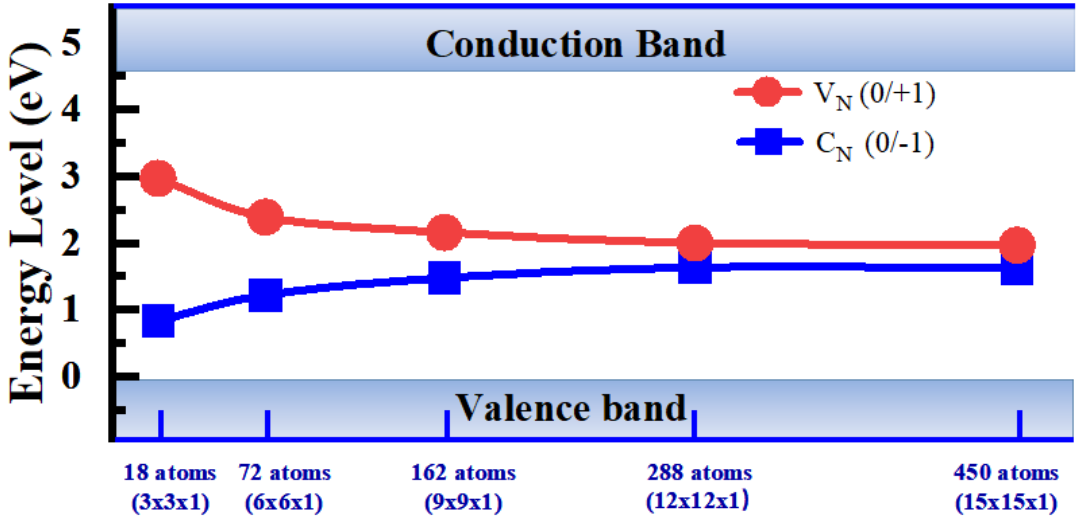

Figure 3. Defect IEs of $V_N^+$ and $C_N^-$ as functions of lateral supercell sizes with the supercell length along the vacuum direction fixed at 15 Å.

Note that, our method can be directly used to study defect IEs in any LD semiconductors. Using our method, we have studied defect properties of several typical 2D semiconductors including BN, black-phosphorous, and MoS$_2$ monolayers. The calculated defect IEs are list in Table I, in comparison with available results of other methods based on the conventional jellium model. Compared to other methods, our method is free of any artificial and tunable parameters like dielectric functions. More importantly, to determine defect IEs, we just need to use one supercell with a moderate size to calculate the total energies of neutral and charged defective supercells. By avoiding possible extrapolations, our method keeps the computational cost lowest. Besides, our method can deal with both donors and acceptors. For example, we have considered group-IVA elements (C, Si, Ge, and Sn) and group-VIA element (O, S, Se, and Te) substituting P in the black phosphorous monolayer. For donors, the defect IEs follow the trend $O_P>S_P>Se_P>Te_P$, because the atomic levels become shallower from O to Te and thus the defects are easy to donate electrons, in agreement with the results of Wang et. al.[37] obtained using the extrapolation method. However, for acceptors, Wang et. al. pointed out that the extrapolation method might fail to give correct defect IEs due to unphysical charge transfer from the materials to the vacuums (see also Fig. S3). In contrast, such charge transfer will not happen in our method and we can get meaningful defect IEs for acceptors. As seen in Table I, the acceptor IEs follow the trend $C_P>Si_P>Ge_P>Sn_P$. This is because these acceptor states are pushed from the valance bands of black phosphorous. From C to Sn, the bond length between the defect atom and phosphorous gets longer. Consequently, the repulsion between defect states and valance bands gets weaker, making the acceptor IEs smaller.

We notice that, our calculated defect IEs are systematically smaller than those using the methods based on the jellium model. This can be simply understood from the Coulomb interactions between the background charge and the charge in the material regions. In our method, the background charge density is physically meaningful band edge charge density, which is distributed only within the materials. On the other hand, the background charge density in the jellium model is distributed in the whole space. As a result, the Coulomb interaction between the background and the material charge is stronger in our method. As the background charge has an opposite sign to the net charge in the material regions, the total energies of charged defective supercells in our method are smaller, giving smaller defect IEs and indicating previous works might have systematically overestimated defect IEs. We also compared our results with a recent work by Xiao et. al[45], in which they proposed a model by transferring charge from defect state to real host band edge states to simulate the charged defect. Using the same unrelaxed structures, our results agree very well with theirs (see the Supplemental Materials).

Table I. Calculated defect IEs in some typical 2D semiconductors using our charge correction method in comparisons with available references using conventional methods based on the jellium model.

| Systems | Defect | Charge state | charge correction method (eV) | References (eV) |
|---|---|---|---|---|
| 2D-BN | $V_B$ | -1 | 0.97 | 1.44[42] |
| | $V_N$ | +1 | 2.01 | 2.50[42] |
| | $C_B$ | +1 | 1.58 | 2.03[42]/2.24[46] |
| | $C_N$ | -1 | 1.39 | 1.86[42]/2.0[46] |
| quasi-2D-BP | $O_P$ | +1 | 0.79 | >0.91[37] |
| | $S_P$ | +1 | 0.50 | 0.74[37] |
| | $Se_P$ | +1 | 0.44 | 0.69[37] |
| | $Te_P$ | +1 | 0.35 | 0.67[37] |
| | $C_P$ | -1 | 0.30 | |
| | $Si_P$ | -1 | 0.25 | |
| | $Ge_P$ | -1 | 0.17 | |
| | $Sn_P$ | -1 | 0.16 | |
| quasi-2D-MoS$_2$ | $V_S$ | -1 | 1.28 | 1.40[47] |
| | $V_{Mo}$ | -1 | 0.64 | 0.85[47] |
| | $Re_{Mo}$ | +1 | 0.12 | 0.22[47] |
| | $F_S$ | +1 | 0.35 | 0.65[47] |

**Conclusions.** In summary, we have proposed a charge correction method based on the jellium model to deal with charged defects by replacing the unphysical jellium background charge with the band edge charge. We have justified the physical meaning of our method and demonstrated it can eliminate the divergence issue in the conventional methods to determine defect IEs. By studying defects in typical 2D materials, we have shown that other conventional methods based on the jellium model might have systematically overestimated defect IEs for LD semiconductors due to the overestimation of Coulomb energy between the background and the material charge. Free of any tunable model parameters, our charge correction method provides a physically meaningful, quantitively accurate and technically simple method to deal with charged defects adapted not only to LD but also to 3D semiconductors.

## Calculation Methods

We implemented the method in the Quantum Espresso code[48]. As the code adopts a self-consistent procedure for electronic relaxation, the charge density correction should be done in each electronic iteration within the conventional jellium model framework. In this case, the corrected background charge density, that is, the charge density of the band edge state, enters the self-consistent loop, which guarantees its contribution to the potential energy and the total energy. The flow chart of our charge corrected method is given in Fig. S2. Using our method, we have studied defect properties of monolayer BN, MoS$_2$, and black phosphorene. The norm-conserving pseudopotentials within the Perdew-Burke-Ernzerhof (PBE) framework [49-51] are used to treat the valence electrons. For the Brillouin zone integrals in the reciprocal space, single Gamma point is used for all calculations for simplicity. The kinetic energy cutoff energy of the plane wave basis is 90 Ry, and the total energy threshold for convergence is $10^{-8}$ Ry. All atoms are relaxed until the Hellman-Feynman forces on individual atoms are less than $10^{-4}$ Ry/Bohr. The VBM and CBM states are implicitly aligned to the levels in defective supercells using vacuum levels. To determine the defect formation energies, we calculated the total energy $E(\alpha, q)$ for a supercell containing the relaxed defect $\alpha$ in its charge state $q$. We also calculated the total energy $E(host)$ for the same supercell in the absence of the defect, as well as the total energies of elemental solids or gases at their stable phases. The defect formation energy $\Delta H_f(\alpha, q)$ as a function of the electron Fermi energy $E_F$ and the atomic chemical potentials $\mu_i$ is given by [52]:

$$\Delta H_f(\alpha, q) = \Delta E(\alpha, q) + \sum n_i \mu_i + q E_F , (1)$$

where $\Delta E(\alpha, q) = E(\alpha, q) - E(host) + n_i E(i) + q E_{VBM}$, $E_F$ is referenced to the VBM of perfect systems, and $\mu_i$ is the chemical potential of constituent $i$ referenced to elemental solid or gas with energy $E(i)$. The $n_i$ are the numbers of atoms taken out of the supercell to form the defects, and $q$ is the number of electrons transferred from the supercell to the Fermi reservoirs in forming the defect cell. Here in the followings, the defect formation energies are given by setting $\mu_i$ and $E_F$ as zeros unless otherwise specified.

## SUPPORTING INFORMATION

Diagrams to show electron occupations for the donor, the sketch of self-consistent calculation flow for charged defect total energy calculation, the failure of the conventional jellium model for acceptor defects, calculated defect IEs using our charge corrected jellium model in comparisons with a recent work by Xiao et. al[45].

## AUTHOR INFORMATION
## CORRESPONDING AUTHOR
E-mail: xggong@fudan.edu.cn

E-mail: jhyang04@fudan.edu.cn

**Notes:**

The authors declare no competing financial interest.

## ACKNOWLEDGMENT

This work was supported in part by NSFC (Grants No. 11974078), Fudan Start-up funding (JIH1512034) and Shanghai Sailing Program (19YF1403100). Calculations are performed at the Supercomputer Center of Fudan University.